\documentclass{aa}
\usepackage[totalwidth=181mm,totalheight=251mm]{geometry}
\usepackage{ulem}
\usepackage{graphicx}

\def\ecs{erg~cm$^{-2}$s$^{-1}$}
\def\lum{erg~s$^{-1}$}
\renewcommand{\tabcolsep}{1.5mm}  

\begin{document}

\title{Six new candidate ultracompact X-ray binaries}

\titlerunning{Six new candidate ultracompact X-ray binaries} 
\authorrunning{J.J.M. in 't Zand, P.G. Jonker \& C.B. Markwardt}

\author{
J.J.M.~in~'t~Zand\inst{1,2},
P.G. Jonker\inst{1,2,3} \&
C.B. Markwardt\inst{4,5}
%\inst{3,4}
}

% \offprints{J.J.M. in 't Zand, email {\tt jeanz@sron.nl}}

\institute{     SRON Netherlands Institute for Space Research, Sorbonnelaan 2,
                3584 CA Utrecht, the Netherlands 
	 \and
                Astronomical Institute, Utrecht University, P.O. Box 80000,
                3508 TA Utrecht, the Netherlands
         \and
                Harvard-Smithsonian Center for Astrophysics, 60 Garden Street,
                Cambridge, MA 02138, U.S.A.
	 \and
	        Department of Astronomy, University of Maryland,
                College Park, MD 20742, U.S.A.
         \and
                Astroparticle Physics Laboratory, Mail Code 661, NASA
                Goddard Space Flight Center, Greenbelt, MD 20771, U.S.A.
	}

\date{Manuscript accepted on January 8, 2007}

\abstract{Ultracompact X-ray binaries (UCXBs) appear able to sustain
  accretion onto the compact accretor at rates lower than in wider
  X-ray binaries. This may be understood by the smaller accretion
  disks in UCXBs: a lower X-ray luminosity suffices to keep a disk
  completely ionized through irradiation and, thus, keep the viscosity
  at a sufficiently high level to allow effective transport of matter
  to the compact object. We employ this distinguishing factor on data
  from RXTE and BeppoSAX to identify six new candidate UCXBs, thus
  increasing the population by one quarter. The candidates are drawn
  from the population of persistently accreting and type-I X-ray
  bursting low-mass X-ray binaries. The X-ray bursts establish the
  low-mass X-ray binary nature and provide a handle on the accretion
  rate.  We find that the low accretion rates are supported by the
  long burst recurrence times and the hard X-ray spectra of the
  persistent emission as derived from the 2nd INTEGRAL catalog of soft
  $\gamma$-ray sources. We discuss the peculiar light curves of some
  new UCXB candidates. }

\maketitle 

\section{Introduction}
\label{intro}

Ultracompact X-ray binaries (UCXBs) are binaries with orbital periods
shorter than $P_{\rm orb}\approx$1~hr in which a neutron star or black
hole accretes matter from a companion star. They are a subset of the
low-mass X-ray binaries (LMXBs). The short orbital period implies such
a small Roche lobe that the donor star must be hydrogen-poor (Nelson
et al. 1986; Savonije et al. 1986). This has 2 interesting
implications: 1) they present interesting laboratories to study
accretion and thermonuclear burning on neutron star surfaces under
hydrogen-poor conditions; and 2) the donor's low-mass (i.e., a few
hundredths of a solar mass) inner core is stripped from its outer
layers yielding an unimpeded view of the ashes of the stellar nuclear
burning which can be studied when dumped on a companion neutron star
or black hole (e.g., Deloye \& Bildsten 2003).

Finding UCXBs is difficult, because measuring $P_{\rm orb}$ is
difficult in LMXBs. Only for eight LMXBs has $P_{\rm orb}$ been
measured with certainty to be in the ultracompact regime, see
Table~\ref{tabucxb}. We call these {\it certain} UCXBs. The shortest
is 11~min, two systems reside at 21 (or 13) and 23 minutes and the
remaining five systems are between 40 and 50 min. Furthermore, there
are four LMXBs for which tentative measurements of $P_{\rm orb}$
exist.  The three UCXBs in globular clusters are all among the
shortest binaries. UCXBs are five times more common in globular
clusters (where 50\% of all measured $P_{\rm orb}$'s are less than
1~hr) than in the Galactic field (Deutsch et al. 2000; Verbunt \&
Lewin 2006).  These differences must be related to different
evolutionary scenarios between cluster and field systems. The large
probability of stellar encounters in globular clusters is an appealing
explanation (Fabian, Pringle \& Rees 1975).

\begin{table}
\begin{center}
\caption[]{List of 27 (candidate) UCXBs (adapted from Nelemans \&
Jonker 2006a), including 6 cases proposed here on the basis of very
low $L_{\rm X}$. We leave out cases identified through the diagnostic
of the X-ray continuum parameter values (Sidoli et al. 2001), for
instance EXO 1745-248 in Terzan 5 (Heinke et al. 2003), because that
diagnostic is not always consistent with others (e.g., Verbunt \&
Lewin 2006).
\label{tabucxb}}
\begin{tabular}{lllll}
\hline\hline
Name                       & (1)   & (2) & (3) & $P_{\rm orb}$ \\
                           &       &     &     & (min)         \\
\hline
\multicolumn{5}{c}{\it certain UCXBs} \\
\hline
XTE J0929-314              &  pp   & T & M   & 44$^{\rm a}$  \\
4U 1626-67                 &  pp   & P & P   & 42$^{\rm b}$  \\
XTE J1751-305              &  pp   & T & M   & 42$^{\rm c}$  \\
XTE J1807-294              &  pp   & T & M   & 40$^{\rm d}$  \\
4U 1820-303 (in NGC 6624)  &  px   & P & B   & 11$^{\rm e}$  \\
4U 1850-087 (in NGC 6712)  &  po   & P & B   & 21 or 13$^{\rm f}$  \\
4U 1915-05                 &  px   & P & B,D & 50$^{\rm g}$  \\
M15 X-2     (in M15)       &  po   & P & B   & 23$^{\rm h}$  \\
\hline
\multicolumn{5}{c}{\it candidate UCXBs with tentative orbital periods} \\
\hline
4U 0614+091                &  po,r & P & B   & 50$^{\rm i}$ \\
4U 1543-624                &  po   & P &     & 18$^{\rm j}$  \\
H 1825-331 (in NGC 6652)   &  po   & P & B   & 55$^{\rm k}$ \\
NGC 6652 B  (in NGC 6652)  &  po   & Q &     & 44$^{\rm k}$ \\
\hline
\multicolumn{5}{c}{\it candidate UCXBs with low optical to X-ray flux} \\
\hline
4U 0513-40   (in NGC 1851) &  r$^{\rm l}$    & P & B   &               \\
2S 0918-549                &  r$^{\rm l}$    & P & B   &               \\
1A 1246-588            	   &  r$^{\rm m}$,x  & P & B   &               \\
4U 1812-12             	   &  r$^{\rm m}$,x  & P & B   &               \\
4U 1822-000            	   &  r$^{\rm l}$    & P &     &               \\
4U 1905+000            	   &  r$^{\rm n}$    & T & B   &               \\
$\omega$ Cen qLMXB     	   &  r$^{\rm o}$    & Q &     &               \\
\hline
\multicolumn{5}{c}{\it candidate UCXBs based on method here discussed} \\
\hline
SAX J1712.6-3739           &  x    & P & B   &               \\
1RXS J170854.4-321857      & x$^{\rm p}$ &P&B&               \\
1RXS J171824.2-402934      & x$^{\rm p}$ &P&B&               \\
4U 1722-30 (in Terzan 2)   &  x    & P & B   &               \\
1RXS J172525.2-325717      &  x    & P & B   &               \\
SLX 1735-269               &  x    & P & B   &               \\
SLX 1737-282               &  x    & P & B   &               \\
SLX 1744-299               &  x    & P & B   &               \\
\hline\hline
\end{tabular}
\end{center}

\noindent
(1) Type of argument for ultracompact identification:
    r = $L_{\rm x}/L_{\rm opt}$ argument,
    p = period measurement (pp=pulsar, px=dips/eclipse, po=optical modulation),
    x = persistent burster with low L;
(2) Type of accretion: P = persistent, T = transient, Q = quiescent
    (never seen to be luminous);
(3) Type of source: P = pulsar, M = accretion-powered millisecond
    pulsar, B = burster, D = eclipser and/or dipper;
$^{\rm a}$Galloway et al. (2002);
$^{\rm b}$Middleditch et al. 1981;
$^{\rm c}$Markwardt et al. (2002);
$^{\rm d}$Markwardt et al.  2003; 
$^{\rm e}$Stella et al. 1987;
$^{\rm f}$Homer et al. 1996;
$^{\rm g}$White \& Swank 1982;
$^{\rm h}$Dieball et al. 2005;
$^{\rm i}$O'Brien et al. 2005;
$^{\rm j}$Wang \& Chakrabarty (2004);
$^{\rm k}$Heinke et al. 2001;
$^{\rm l}$Juett et al. 2001;
$^{\rm m}$Bassa et al. 2006;
$^{\rm n}$Jonker et al. 2006;
$^{\rm o}$Haggard et al. 2004;
$^{\rm p}$in 't Zand et al. 2005a.
\end{table}

Identification of UCXBs is most directly done through measuring
$P_{\rm orb}$.  There are three methods for this measurement: 1)
through timing of Doppler-delayed pulses if the accretor is a pulsar
(four detections of certain UCXBs resulted from this, most notably
three transient accretion-powered millisecond pulsars); 2) through
measuring periodic eclipses, dips or modulations of X-rays if the
inclination angle is high enough (two detections); and 3) through
measurements of periodic optical modulations (three detections)
resulting possibly either from heating one side of the donor by X-ray
irradiation or from the superhump phenomenon that is predicted for
mass ratios far from 1 (like expected in many UCXBs).

There are two indirect methods to identify UCXBs without measuring
$P_{\rm orb}$.  Both depend on the notion that in an UCXB the
accretion disk must be relatively small. The first indirect method
concerns the ratio of optical to X-ray flux. At the same X-ray flux,
$M_V$ is about 4 mag fainter for UCXBs than for non-ultracompact LMXBs
(van Paradijs \& McClintock 1994). Seven UCXBs without measured
$P_{\rm orb}$'s have so far been identified following this method
(e.g., Juett et al. 2001). The second indirect method concerns the
critical accretion rate below which a system becomes transient and is
employed in this paper to identify six new UCXBs. We present the
principle of the method in Sect.~\ref{method}, the tools in Sect.
\ref{bursts}, the data in Sect. \ref{data}, the results in Sect.
\ref{results}, corroborative evidence in Sect. \ref{integral} and a
discussion of the results in Sect. \ref{discussion}.

We note that there are additional promising diagnostics for a UCXB
nature and they are all based on spectral data. The most direct method
involves measuring the composition of the donor through spectral lines
and edges in the optical accretion disk spectrum (e.g., Nelemans et
al. 2004; Werner et al. 2006), in particular the absence of hydrogen
lines. X-ray spectroscopy was initially also promising: Juett et
al. (2001) and Juett \& Chakrabarty (2003) interpreted unusually high
Ne/O abundance ratios as due to a UCXB nature. However, later these
ratios turned out to be variable, thus weakening this interpretation
(Juett \& Chakrabarty 2005). Finally, Sidoli et al. (2001) made a
comparison of the X-ray continua spectra of bright globular cluster
LMXBs and noted a dichotemy between ultracompact and non-ultracompact
cases; the parameters of the disk black body components in
ultracompact cases appear to be physically more realistic and
consistent with Comptonization components than those in
non-ultracompact cases.

\section{Principle: a critical mass transfer rate below which accretion
becomes transient}
\label{method}

Cataclysmic variables (CVs; Smak 1983; Osaki 1996) and LMXBs (White et
al. 1983) remain in a persistently accreting state if the mass
transfer rate from the donor star is above a certain critical value
$\dot{M}_{\rm crit}$. The disk instability model (e.g., Osaki 1974;
Hoshi 1979; Osaki 1996; Lasota 2001) provides a natural explanation
for this.  Below the critical value the disk is unstable and steady
accretion is impossible. Since $\dot M_{\rm crit}$ is strongly
increasing with radius it is its value at the outer disk radius that
determines its stability. Therefore the value of $\dot{M}_{\rm crit}$
is a strong function of the orbital period and mass (Smak
1983). Furthermore, there is a distinction between CVs and LMXBs
because X-ray irradiation is an important effect in the ionization
balance of LMXB accretion disks while it is not in CVs (van Paradijs
1996).

Evaluating $\dot{M}_{\rm crit}$ is difficult, because of the various
uncertainties in the nature of the viscosity and the geometry of the
irradiation (related to the questions whether the source is point like
or extended with respect to the disk and whether the disk is warped).
Dubus et al. (1999) and Lasota (2001) derive\footnote{This is Eq. (36)
in Lasota (2001) after correcting the numerator of Eq. (34), which was
used in deriving Eq. (36), from $M_2$ to $M_1$ (Lasota, priv. comm.)}
\begin{eqnarray}
\dot{M}_{\rm crit} & = & 5.3 \times 10^{-11}C\hspace{3mm} M_1^{0.3}
P_{\rm orb}^{1.4} \hspace{3mm} M_\odot\;{\rm yr}^{-1}
\end{eqnarray}
with accretor mass $M_1$ (in $M_\odot$) and orbital period $P_{\rm
orb}$ (in hr). $C\approx1$ represents uncertainty\footnote{We employ
another expression for $C$ than Dubus et al. (1999) and Lasota
(2001)}. For $P_{\rm orb}<1$~hr and $M_1=1.4$~~M$_\odot$,
$\dot{M}_{\rm crit}\la 6\times10^{-11}$~M$_\odot\;{\rm yr}^{-1}$. This
value is for solar composition.  Menou et al. (2002) calculated values
for other compositions, although for the non-irradiated case. For C/O
disks the results are equal to the solar composition case within a few
tens of percents.  For pure He or O disks, the value is 6 times
larger. Therefore, if we naively apply this factor to the irradiated
case, $\dot{M}_{\rm crit}$ is approximately 1\% or less of the
Eddington limit for a hydrogen-poor/helium-rich photosphere of a
bursting 1.4~M$_\odot$ neutron star. Herein lies the principle of our
method: a bursting LMXB may be tentatively identified as a UCXB if it
is persistent at accretion luminosities below $\approx1$\% of
Eddington.

\begin{figure*}[!t]
\centering
\includegraphics[width=2.0\columnwidth]{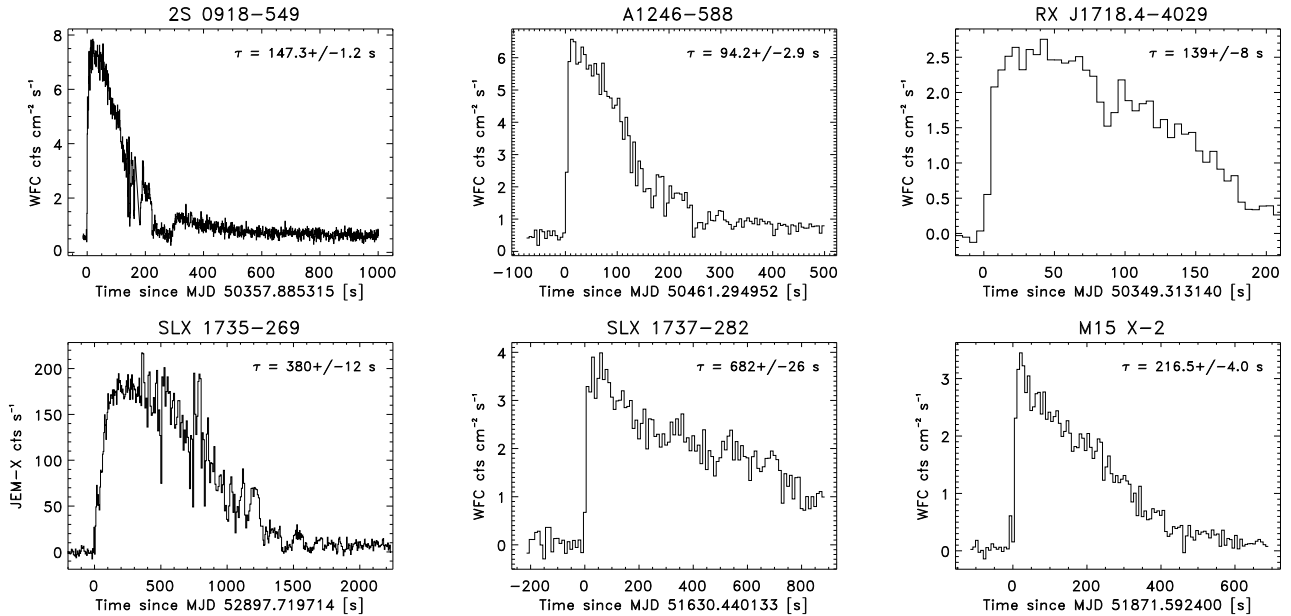}
\caption{A compilation of 2-30 keV time profiles of longest bursts
from 5 (candidate) UCXB candidates with BeppoSAX/WFC and 1 with
INTEGRAL/JEM-X (Molkov et al. 2005). Indicated are the e-folding decay
times resulting from a fit to the data beyond the peak flux and
excluding clear deviations from an exponential as for instance in 2S
0918-549. All these bursts start with a radius-expansion
phase.\label{figbursts}}
\end{figure*}

It is difficult to measure $\dot{M}$, because the uncertainties are
large in translating observed flux to bolometric luminosity (i.e.,
uncertainties in bolometric and anisotropy corrections) and in
translating bolometric luminosity to mass accretion rate (i.e.,
uncertainty in radiation efficiency).  Combined with the uncertainty
in $\dot{M}_{\rm crit}$ this makes a sensible test between both
accretion rates impossible.  Nevertheless, there is merit in the
principle. It turns out that if we rank persistently accreting LMXBs
according to their ratios of estimated bolometric flux to Eddington
flux, all UCXBs with known (tentative) $P_{\rm orb}$'s except
one populate the lowest $\dot{M}$ regime.

\section{Tool: the X-ray burst phenomenon in UCXBs}
\label{bursts}

We apply the principle just described to those persistently accreting
LMXBs that exhibit type-I X-ray bursts (or shortly, 'X-ray bursts')
which provides an easy estimate of the Eddington flux.  X-ray bursts
result from thermonuclear flashes of hydrogen or helium in the freshly
accreted layers of neutron stars (for reviews, see e.g.  Lewin et
al. 1993 and Strohmayer \& Bildsten 2006). For a few seconds to
minutes the flashes heat up the photoshere to few-keV temperatures and
the resulting X-ray spectra are dominated by easy-to-analyze black
body shapes. The peak fluxes are close to the Eddington limit and for
many luminous bursts even equal to it.

Fortunately, many UCXBs exhibit X-ray bursts, although it is not
really understood why.  Four of the eight certain UCXBs exhibit type-I
X-ray bursts, as do seven of the eleven candidate UCXBs. Since the
hydrogen abundance in UCXBs is neglible, the bursts must result from
flashes of helium being accreted from the companion star.  A
conclusive confirmation of this argument would be the detection of
helium in the optical spectrum. Within the group of UCXBs that burst
and have (tentative) $P_{\rm orb}$ values, this confirmation is there
for only one case (XB 1916-05; Nelemans et al. 2006b) showing a He-II
emission line at 4690~\AA, albeit comparatively weak, while it is
significantly absent in another (4U 0614+091; Nelemans et al. 2004 and
Werner et al. 2006). The remaining cases lack good enough optical
spectra for definite verification although He lines should have been
detected in some cases if they would have had similar equivalent
widths as in wide-orbit LMXBs. The lack of He lines is one of the
reasons why the presence of X-ray bursts is not understood (e.g.,
Nelemans et al. 2006b). Disregarding that, we use the presence to our
advantage.

Often the X-ray bursts from both certain as well as candidate UCXBs
are rather long. They sometimes are the longest after the superbursts
(up to 40 min, e.g. in 't Zand et al. 2005b; for a few examples see
Fig.~\ref{figbursts}). Cumming et al. (2006), therefore, coined the
term ' intermediately long X-ray bursts'. In general, burst duration
is determined by the thickness and hydrogen abundance of the flash
layer immediately prior to the flash. Both of these depend foremost on
the mass accretion rate.  Most bursts are short with a duration of
approximately 10~s and they occur in a very specific accretion rate
regime (Fujimoto et al. 1981; Bildsten 1998). The reason that there
are so many of them is that many LMXB reside in this accretion regime
and the associated burst recurrence time is short (i.e., of order a
few hours). The relevant accretion regime is roughly between 1 and
10\% of Eddington (for details, see Bildsten 1998 and references
therein). The conditions in this regime are such that between flashes
the freshly accreted hydrogen is stably burned at a rate as it is
accreted, thus developing a pure helium layer. When the helium layer
is thick enough, the pressure at the bottom surpasses ignition
conditions for the runaway 3$\alpha$ process. The flash ends within
1~s. The burst duration is a direct measure of the layer cooling time
(between a few seconds and a few tens of seconds) and thus depends on
the layer thickness. Outside this accretion regime bursts are
longer. If the accreted matter is hydrogen rich, this is due to slow
beta decay of the products of rapid proton capture by the ashes of the
initial helium or hydrogen flash, prolonging the nuclear energy
generation by a few minutes. If the accreted matter is hydrogen poor,
a long burst is due a low layer temperature if the long-term average
accretion rate is so low that heating by pycnonuclear reactions in the
crust is significantly reduced (see in 't Zand et al. 2005a and
Cumming et al. 2006)\footnote{Crustal heating reacts to mass accretion
rate on time scales of months and, thus, is independent of short term
variability}. Only long bursts of the latter kind can occur in UCXBs.

It is expected that X-ray bursts of the latter kind easily reach the
Eddington limit resulting in photospheric radius expansion (PRE),
because the triple-$\alpha$ reaction rates are fast and the amount of
helium large.  Also, the PRE phase may last long because of the large
helium layer thickness (of order 1 min; c.f., Cumming et
al. 2006). The long mixed hydrogen/helium flashes do not last longer
than a few hundred seconds and may have irregularly shaped decay
phases due to the various waiting points in the rp-process reaction
chain (e.g., Schatz et al. 2001; Woosley et al. 2004). The
characteristics of the various long X-ray bursts suggest that a burst
with a duration in excess of a few minutes and exhibiting long PRE is
a unique diagnostic of a UCXB. This appears to be consistent with the
fact that such bursts have never, as far as we know, been detected
from non-ultracompact systems.

Nevertheless, we do not employ such bursts as a prime diagnostic.
Recent modeling of the effect of sedimentation on burst ignition by
Peng et al.  (2006) shows that for certain low mass accretion rates
(below 1\% of the Eddington limit) of {\it hydrogen-rich} material
(ergo, in a non-ultracompact binary) there may exist a burst regime
with flashes of pure hydrogen in a layer that is too thin to ignite
helium. These flashes produce helium through the CNO cycle. A thick
helium layer grows which may result in a similarly long burst as
described above for a hydrogen-poor ultracompact case at low accretion
rate. There appear two issues concerning the viability of this burst
regime: 1) as Peng et al. (2006) discuss, for mass accretion rates of
order 1\% of the Eddington limit the pure helium layer would grow to
10$^{11}$ g~cm$^{-2}$ before ignition would occur which would result
in burst durations comparable to those of short superbursts which has
never been observed; 2) as we have discussed in Sect.~\ref{method},
the low accretion rates needed may not be possible in non-ultracompact
cases because the disk instability would turn off the
accretion. Resolution of these issues is desirable but difficult
because of the nature of the objects involved: quite faint objects
with predicted superburst-like flares every few years (note that these
are inconsistent with 'burst-only' sources; see Cornelisse et
al. 2002a and 2002b).

\section{Data}
\label{data}

\subsection{Average persistent fluxes of persistent X-ray bursters}

\begin{table*}[!t]
\begin{center}
\caption[]{40 persistent X-ray bursters, in order of average to
Eddington flux ratio. The division lines are at 2 and
10\%.\label{burstlist}} \vspace{-0mm}
\begin{tabular}{lrrrrcrrrrrr}
\hline\hline
Source name & \multicolumn{2}{|c|}{ASM} & \multicolumn{2}{|c|}{PCA} & \multicolumn{1}{|c|}{ASM/} & \multicolumn{1}{|c|}{Other} & \multicolumn{1}{|c|}{Burst} &  \multicolumn{1}{|c|}{Ratio} & \multicolumn{1}{c|}{Burst rec.} & \multicolumn{1}{c}{Previously} & \multicolumn{1}{|c}{$P_{\rm orb}$} \\
            & \multicolumn{2}{|c|}{   } & \multicolumn{2}{|c|}{   } & \multicolumn{1}{|c|}{PCA}  & \multicolumn{1}{|c|}{     } & \multicolumn{1}{|c|}{peak} &\multicolumn{1}{|c|}{\%} &\multicolumn{1}{|c|}{time (hr)}& \multicolumn{1}{|c}{identified}& \multicolumn{1}{|c}{(hr)} \\
            & \multicolumn{2}{|c|}{   } & \multicolumn{2}{|c|}{   } & \multicolumn{1}{|c|}{} & \multicolumn{1}{|c|}{     } & \multicolumn{1}{|c|}{flux} &\multicolumn{1}{|c|}{Edd.} &\multicolumn{1}{|c|}{}& \multicolumn{1}{|c}{UCXB?} & \multicolumn{1}{|c}{}\\
            & \multicolumn{1}{|c}{(1)} & \multicolumn{1}{c|}{(2)} & \multicolumn{1}{|c}{(1)} & \multicolumn{1}{c|}{(2)} & \multicolumn{1}{|c|}{} & \multicolumn{1}{|c|}{(3)} & \multicolumn{1}{|c|}{(3,4)} & (5) & \multicolumn{1}{|c|}{(6)} & \multicolumn{1}{|c}{}& \multicolumn{1}{|c}{(7)}\\
\hline
1RXS J171824.2$^{(8)}$&  0.402(9)  &   3.1 & 	-	&   - 	&      & 0.1$^{\rm a}$&	 \underline{390}$^{\rm b}$ &  0.03 & 438--8254    &  & \\
SLX 1737-282          &  -         &   -   &   47.8(7)	&   2.4	&      & 2.3$^{\rm c}$&  \underline{600}$^{\rm c}$ &   0.4 & 412--7778    &  & \\
2S 0918-549           & 0.576(5)   &   4.5 & 	-	&   - 	&      & 6.0$^{\rm d}$&	\underline{1000}$^{\rm d}$ &   0.5 & 202--853     & cand. UCXB & \\
1A 1246-588           & 0.618(6)   &   4.8 & 	-	&   - 	&      &	      &	 \underline{900}	   &   0.5 & $278\pm139$  & cand. UCXB & \\
SAX J1712.6-3739      &  0.868(9)  &   6.7 &   50.7(7)	&   2.6	& 2.58 &	      &	 \underline{510}$^{\rm e}$ &   0.5 & 345--6507    &  & \\
4U 1812-12            &  1.328(7)  &  10.3 &  166.1(8)	&   8.5	& 1.21 &10.9$^{\rm f}$&	\underline{1600}           &   0.5 & $80.2\pm18.9$& cand. UCXB &\\
4U 1850-087           &  0.606(5)  &   4.7 &  52.1(2)	&   2.7	& 1.74 &11.9$^{\rm g}$&	 \underline{600}$^{\rm h}$ &   0.5 & $>1584$       & UCXB & 0.3 \\
\multicolumn{2}{l}{1RXS J172525.2-325717 \hfill -} &   -   &   24.9(2)	&   1.3	&      & &  230$^{\rm ac}$ &   0.6 &             &  & \\
4U 0614+091           & 3.111(5)   &  24.1 & 	-	&   - 	&      &	      &	\underline{3000}$^{\rm i}$ &   0.8 & 168--3175    & cand. UCXB & 0.8 \\
SLX 1735-269          &  1.25(1)   &   9.7 &  107.0(3)	&   5.5	& 1.76 &	      &	 \underline{577}$^{\rm j}$ &   1.0 & 387--7301   &  & \\
EXO 0748-676          & 0.668(5)   &   5.2 & 	-	&   - 	&      &	      &	 \underline{520}$^{\rm k}$ &   1.0 & $5.1\pm0.4$ & & 3.8 \\ % eclipser
4U 1915-05            &  1.001(5)  &   7.7 &   -	&   - 	&      &	      &	 646$^{\rm l}$             &   1.1 & 31$\pm$11    & UCXB & 0.8 \\ %dipper
H 1825-331            &  0.665(9)  &   5.1 &  70.5(5)	&   3.6	& 1.42 & 4.3$^{\rm m}$&	 \underline{297}$^{\rm n}$ &   1.2 & $27.8\pm7.4$   & cand. UCXB & \\
M15 X-2$^{(9)}$       &  0.565(3)  &   4.4 &   -	&   - 	&      & 3.8$^{\rm g}$&	 \underline{375}$^{\rm p}$ &   1.2 & 37--984      & UCXB & 0.4 \\
XTE J1710-281         &  0.425(12) &   3.3 &   23.04(5)	&   1.2	& 2.75 &	      &	 \underline{92}$^{\rm n}$  &   1.3 &              & & 3.9 \\ % eclipser           ***
\multicolumn{2}{l}{1RXS J170854.4-321857 \hfill -} &-& -	&   - 	&      & 2.4$^{\rm a}$&	 \underline{154}$^{\rm a}$ &   1.5 &  101--1904            & \\
4U 1722-30           &  2.06(1)   &  15.9 &  246.4(4)	&  12.6	& 1.26 & 18$^{\rm g}$ &	 \underline{708}$^{\rm o}$ &   1.8 & $57\pm12$&  & \\
4U 0513-40            & 0.411(4)   &   3.2 & 	-       &   -   &      &              &	 \uwave{170}$^{\rm n}$     &   1.9 & $49\pm14$    & cand. UCXB & \\
SLX 1744-299$^{(10)}$  & 0.989(8)   &   7.7 &    163(5)  &   8.3 & 0.92 & 12$^{\rm r}$ & 420$^{\rm q}$              &   1.9 & 188--793    & & \\
\hline
4U 1746-37            &  2.306(8)  &  17.8 &  318.6(5)	&  16.3	& 1.09 &	      &	 \underline{630}$^{\rm n}$ &   2.6 &              & & 5.7 \\
A 1742-294            &  -         &   -   &  213.2(19)	&  10.9	&      &	      &	 \underline{401}$^{\rm n}$ &   2.7 & $6.1\pm0.4$  & & \\
4U 1702-429           &  3.191(8)  &  24.7 &  429.0(11)	&  21.9	& 1.13 &	      &	 \underline{810}$^{\rm n}$ &   2.7 & $11.4\pm1.0$ & & \\                      %***
XTE J1759-220         &  0.556(10) &   4.3 &   31.45(8)	&   1.6	& 2.69 &	      &	  51$^{\rm n}$             &   3.1 &              & & 1-3 \\ % dipper
SLX 1744-300$^{(10)}$  &  0.534(4)  &   4.1 &   88(3)    &   4.5 & 0.92 & 6$^{\rm r}$  &  190$^{\rm n}$             &  3.2 &  24.7$\pm$6.7         & & \\
4U 1323-62            & 0.598(8)   &   4.6 & 	-	&   - 	&      &	      &	 107$^{\rm n}$             &   4.3 & $39\pm11$   & & 2.9 \\ % dipper             %*** e gone
GX 354-0              &  6.311(8)  &  48.8 & 1031.3(10)	&  52.7	& 0.93 &	      &	\underline{1200}$^{\rm s}$ &   4.4 & $3.2\pm0.2$& & \\
GS 1826-24            &  2.535(10) &  19.6 &  424.8(1)	&  21.9	& 0.89 &              &	 330$^{\rm t}$             &   6.6 & $4.6\pm0.3$& & 2.1? \\
\hline
4U 1636-536           & 10.420(7)  &  80.6 & 	-	&   - 	&      &	      &	 \underline{742}$^{\rm n}$ &  10.9 & $8.9\pm1.0$ & & 3.8 \\                      %*** h gone
4U 1705-440           & 10.857(9)  &  84.0 & 1082.9(16)	&  55.3	& 1.52 &	      &	 \underline{410}$^{\rm n}$ &  13.5 & $16.5\pm1.9$ & & \\
UW Crb                & -          &   -   & 	-	&   - 	&      & 0.4$^{\rm u}$&	2.44$^{\rm u}$             &  16.4  &              & & 1.9 \\
4U 1254-69            & 2.420(5)   &  18.7 &    -       &   -   &      & 14$^{\rm v}$ & \underline{110}$^{w}$      &  17.0 & $44.9\pm8.8$   & & 3.9 \\ % dipper
GX 3+1                & 21.015(12) & 162.5 & 2991.4(12)	& 152.9	& 1.06 &	      &	 \underline{690}$^{\rm x}$ &  22.2 & $21.4\pm2.7$ & & \\
4U 1820-303           & 19.31(1)   & 149.3 & 2754.4(13)	& 140.8	& 1.06 & 	      &	 \underline{570}$^{\rm n}$ &  24.7 & $26.6\pm3.8$ & UCXB & 0.2 \\
4U 1708-40            &  1.910(8)  &  14.8 &  434.8(11)	&  22.2 & 0.67 &	      &	  86$^{\rm aa}$            &  25.8 &              & & \\
4U 1735-44            & 13.234(9)  & 102.3 &    -       &   -   &      &              &  358$^{\rm n}$             &  28.6 & $29.9\pm4.8$ & & 4.6 \\
Ser X-1               & 16.189(6)  & 125.2 &   -	&   - 	&      &	      &	 293$^{\rm n}$             &  42.7 & $75\pm29$    & & \\
Cir X-1               & 13.766(10) & 106.5 & 	-	&   - 	&      &	      &	 204$^{\rm z}$             &  52.2 &              & & 398 \\
GX 13+1               & 22.788(9)  & 176.2 & 3757.5(30)	& 192.0	& 0.92 &	      &	 260$^{\rm y}$             &  73.8 &              & & \\
Cyg X-2               & 35.682(7)  & 275.9 &   -        &   -   &      &              &	 \uwave{154}$^{\rm n}$     & 179.1 &              & & 236 \\                      %***
GX 17+2               & 44.631(10) & 345.1 & 7457.1(42)	& 381.1	& 0.91 &	      &	 \underline{145}$^{\rm ab}$& 262.8 & $105\pm29$ & \\
\hline\hline
\end{tabular}
\end{center}

\vspace{-3mm}
\noindent
(1) Average instrument intensity in c~s$^{-1}$, normalized to 5 PCUs for the PCA; includes data up
    to June 2006; 
(2) Estimated bolometric flux in 10$^{-10}$~\ecs;
(3) Single flux measurements in 10$^{-10}$~\ecs\ derived from a broader bandpass with, thus, more accurate
    bolometric corrections; only values given for faint sources, particularly if they are
    not covered by the PCA bulge scans;
(4) if underlined (with a wave) the flux relates to a (tentative) Eddington-limited case; 
(5) the persistent to burst peak flux ratio, prioritized following PCA flux, 'other'
    flux (note 3) if the ASM data are flat or near expected bias levels (i.e., 0.1 to 0.5 c~s$^{-1}$)
    depending on the sky position, or the ASM flux;
(6) Burst recurrence time from BeppoSAX-WFC archive. Uncertainties and lower limits are for
    68\% confidence from Poisson statistics;
(7) For some references, see Table~\ref{tabucxb};
(8) Full name: 1RXS J171824.2-402934;
(9) The flux of M15 X-2 was scaled from the total M15 flux through the fluxes
    measured for M15 X-2 and AC 211 by White \& Angelini (2001);
(10) SLX 1744-299 and SLX 1744-300 are only 2\farcm8 apart and cannot be separated by the ASM
    nor PCA. The ASM and PCA fluxes of both sources were scaled according to a flux 1.0/2.8 ratio 
    following Mori et al. (2005).;
$^{\mathrm a}$in 't Zand et al. 2005a;
$^{\mathrm b}$Kaptein et al. 2000;
$^{\mathrm c}$in 't Zand et al. 2002a;
$^{\mathrm d}$in 't Zand et al. 2005b;
$^{\mathrm e}$Cocchi et al. 2001;
$^{\mathrm f}$Barret et al. 2003;
$^{\mathrm g}$Sidoli et al. 2001;
$^{\mathrm h}$Hoffman et al. 1980;
$^{\mathrm i}$Kuulkers et al., in prep.;
$^{\mathrm j}$Molkov et al. 2005;
$^{\mathrm k}$Wolff et al. (2005);
$^{\mathrm l}$Smale et al. 1988;
$^{\mathrm m}$ Parmar et al. 2001;
$^{\mathrm n}$Galloway et al. 2006;
$^{\mathrm o}$Molkov et al. 2000;
$^{\mathrm p}$van Paradijs et al. 1990;
$^{\mathrm q}$Pavlinsky et al. 1994 and in 't Zand et al. in prep.;
$^{\mathrm r}$Mori et al. 2005
$^{\mathrm s}$Galloway et al. 2003;
$^{\mathrm t}$ Galloway et al. 2004;
$^{\mathrm u}$Hakala et al. 2005 and Hynes et al. 2004;
$^{\mathrm v}$Iaria et al. 2001;
$^{\mathrm w}$in 't Zand et al. 2003.
$^{\mathrm x}$Kuulkers \& van der Klis 2000;
$^{\mathrm y}$Matsuba et al. 1995;
$^{\mathrm z}$Tennant et al. 1986;
$^{\mathrm {aa}}$Migliari et al. 2003;
$^{\mathrm {ab}}$Kuulkers et al. 2002;
$^{\mathrm {ac}}$Brandt et al. 2006.
\end{table*}

We collected data on all 40 X-ray bursters which, to the best of our
knowledge, are currently active and have been so for longer than five
years, see Table~\ref{burstlist}. It is fair to assume that the
accretion in these systems is persistent because the viscous
timescales in the accretion disk are thought to be only a few months
(e.g., Lasota 2001 and references therein). To estimate the {\it
average} bolometric flux, we employed flux data from two {\it
long-term} monitoring observations. The flux in persistent LMXBs is
known to fluctuate up to an order of magnitude on time scales of up to
hundreds of days. Therefore, a one-time flux measurement may not be a
good representation of the mass transfer rate from the companion
star. Long-term monitoring observations are crucial. We appealed to 1)
public data from the Rossi X-ray Timing Explorer (RXTE) All-Sky
Monitor (ASM) which has been gathering data on most of these sources
since January 1996 (Levine et al. 1996; Wen et al. 2006) in the 2-12
keV band each day for roughly 10 months of the year at a sensitivity
of about $5\times10^{-10}$~\ecs\ per day. Sources which are not
included are either too faint to be detectable, or too near a bright
source to allow accurate flux measurements; 2) the RXTE Proportional
Counter Array (PCA) Galactic Bulge monitoring program which has been
running on the inner 16$^{\rm o}$ of the bulge since February 1999 and
on the inner 52$^{\rm o}$ since May 2004 (Swank \& Markwardt 2001;
Markwardt 2006), measuring the flux of every source twice a week in
the 2-12 keV band for 10 months of each year at a sensitivity of about
$10^{-11}$~\ecs. There are four sources (1RXS J171824.2-402934, SLX
1744-299, SLX 1744-300 and M15 X-2) for which there are no accurate
ASM nor PCA measurements due to faintness or source confusion and we
use results from targeted more sensitive observations. For these, we
lack long enough exposures to assess a good long-term average,
although it should be said that they, apart from during X-ray bursts,
have never been seen in a bright state.

In order to derive from the observed photon flux an estimate of the
bolometric energy flux, the absolute calibrations provided for the two
data sets are employed: the Crab source yields 75 ASM c~s$^{-1}$ and
11350 PCA c~s$^{-1}$ (normalized to 5 Proportional Counter Units). The
Crab spectrum is a power law with a photon index of 2.1 and $N_{\rm
H}=4\times10^{21}$~cm$^{-2}$ (Kirsch et al. 2005). If in fact the
photon index of the actual source differs by 1 or $N_{\rm H}$ is up to
10 times larger, the 2-10 or 2-12 keV energy flux differs by at most
30\%. We assume this to be the uncertainty in translating the 2-10 keV
observed photon flux to 2-10 keV intrinsic energy flux. The next step
is the bolometric correction.  The literature on broad-band spectra of
LMXBs frequently provides absorbed fluxes in 2-10 keV as well as
unabsorbed 0.1-100 keV fluxes. We browsed the literature to obtain
characteristics values for the flux ratio and find $2.9\pm1.4$ to be a
fair representation.  Thus, we have a conversion of 1 ASM c~s$^{-1}$
to 7.7$\times10^{-10}$~\ecs\ and 1 PCA c~s$^{-1}$ to
5.1$\times10^{-12}$~\ecs, with a typical error of a factor of 2.  It
is not really worthwhile to try to assess the spectrum individually
per source from archival data and derive conversion factors from that
because the relevant data are mostly snapshots that are not
representative of the average behavior and because $N_{\rm H}$
measurements usually suffer from such large systematic errors that the
correction for absorption, important below a few keV, is quite
inaccurate.

An additional correction is necessary for the inclination angle: if
the disk is viewed near edge-on, the observable flux reduces
considerably and needs to be corrected by a factor of $\xi_{\rm
p}^{-1}=2|\cos i|$ (Lapidus \& Sunyaev 1985; Fujimoto 1988).
$\xi_{\rm p}$ ranges from 0.5 for $i=0\degr$ to $>2$ for
$i>75\degr$. For $i\ga85\degr$ the accretion disk probably blocks the
view to the NS and no bursts are observable. Without knowledge of $i$,
$\xi_{\rm p}$ probably ranges between 0.5 and 2 for bursters.  An
additional uncertainty is introduced by the efficiency factor $\eta$
with which gravitational energy is transformed to radiation and how
representative the flux is for the mass accretion rate. Combining all
uncertainties probably adds up to a factor of 3 (i.e., the quadratic
sum of 3 factors of 2).

Table~\ref{burstlist} provides the average raw photon count rates for
ASM and PCA, the extrapolated flux values and a comparison between
both values if they are both present (i.e., a few sources are not in
the ASM catalog, and half are not covered in the PCA bulge scan
program). Only 3 out of 19 comparisons deviate by more than a factor
of two. These are also the 3 sources with the lowest fluxes. This
probably results from unaccounted-for bias levels in the ASM fluxes.
For a constant bias level of 0.3 ASM c~s$^{-1}$, which is a reasonable
number for short angular distances to the Galactic center, the ASM/PCA
ratios would remain with a factor of 2 from the value 1.

\subsection{Eddington fluxes}

\begin{figure}[!t]
\centering
\includegraphics[width=0.75\columnwidth]{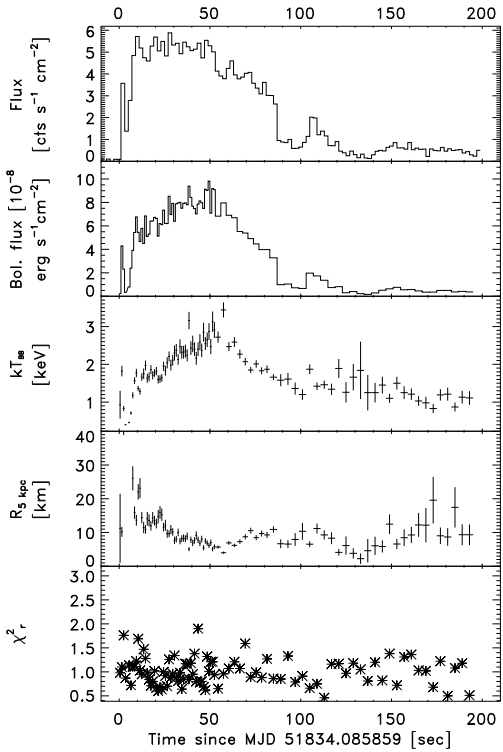}
\caption{Time-resolved spectroscopy of burst from 1A~1246-588
obtained with BeppoSAX-WFC.\label{figbursts2}}
\end{figure}

The highest bolometric burst peak flux observed for any source in the
history of X-ray astronomy provides a lower limit to the Eddington
flux for that source. Fortunately, these maxima often apply to bursts
which experienced photospheric radius expansion (PRE) and the peak
flux is actually equal to the Eddington flux. We searched through the
literature to find the highest peak fluxes. For one case we determined
the burst peak flux ourselves from BeppoSAX-WFC data, see
Fig.~\ref{figbursts2}. The results are listed in
Table~\ref{burstlist}. Fifteen sources did not exhibit any unambiguous
PRE bursts.

For canonical neutron stars with a mass of 1.4~M$_\odot$ and a radius
of 10~km, the Eddington limit is 2.0$\times10^{38}$~\lum\ for a
hydrogen-rich photosphere and 3.5$\times10^{38}$~\lum\ for a
hydrogen-poor photosphere (if the radius expansion is small with
respect to the NS radius, a relativistic correction of a factor of
$\sqrt{1-2GM/Rc^2}$ needs to be applied to these values which equals
0.76 for a canonical NS).  Like the persistent flux, the burst peak
flux also needs to be corrected for inclination angle. Lapidus \&
Sunyaev (1985) and Fujimoto (1988) derive a correction factor of
$\xi_{\rm b}^{-1}=0.5+|\cos i|$. If no orbital modulation is seen in
the flux, $i$ may be presumed to be smaller than 70\degr\ and $\xi_p$
lies between 0.84 and 1.5. Thus, this introduces an uncertainty of
about 30\%.

\subsection{Average mass transfer rates in terms of the Eddington limit}

The burst peak flux represents a measurement of the Eddington limit or
a lower limit to that (if no PRE bursts were detected). The persistent
flux represents the average mass accretion rate which, for persistent
sources, is equal to the mass transfer rate. The ratio thus provides
an indication for the upper limit to the mass transfer rate in terms
of the Eddington limit, with all the uncertainties mentioned
above. The ratio numbers provided in Table~\ref{burstlist} have been
calculated with the best numbers for persistent flux, in other words
the PCA bulge scan numbers take preference over the other 2
numbers. If the ASM data show the source to be reasonably constant
over the years, the fluxes from independent studies ('other' numbers
in Table) take preference over the ASM numbers. The rms uncertainty in
the flux numbers is expected to be close to a factor of 2.

Three groups can be distinguished in the ratio distribution: bursters
with a mass transfer rate higher than $\sim10$\% of Eddington (13
cases including the highest cases GX 17+2 and Cyg X-2); those with a
ratio between $\sim2$ and $\sim10$\% (8 cases); and those with lower
ratios (19 cases). Most importantly: all ten persistent and bursting
LMXBs that have been identified as ultracompact, except 4U 1820-303,
are in the low mass-transfer rate regime.

\subsection{Burst recurrence times}

An interesting parameter is the burst recurrence time since this is
dependent on the mass accretion rate on the neutron star: the faster
new fuel is provided from the donor, the shorter the burst recurrence
time. The recurrence time is not completely inversely proportional to
the accretion rate.  For accretion rates in excess of 1 to 10\% of the
Eddington limit, hydrogen burning is stable and will not give rise to
X-ray bursts (e.g., Fujimoto et al. 1981). Thus, there is a sudden
change of recurrence time at this threshold value.

For the most common bursters that radiate at about 10\% of the
Eddington limit, burst recurrence times are of order a few
hours. Going further above and below that the recurrence time
increases (e.g., Cornelisse et al. 2003).

\begin{figure}[!t]
\centering
\includegraphics[width=0.99\columnwidth,angle=0]{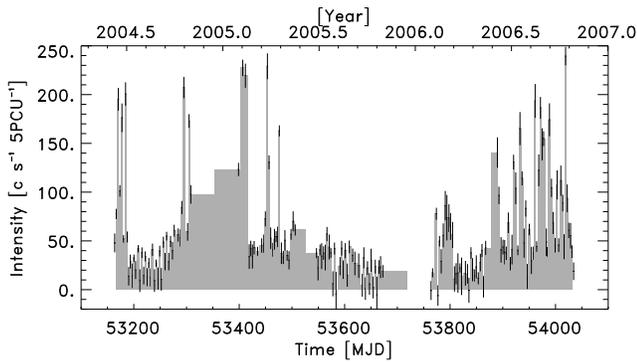}
\caption{The 2-12 keV light curve of SAX~J1712.6-3739 as measured
between May 2004 and November 2006 through the PCA bulge scan program.
\label{fig1712}}
\end{figure}

Probably the most comprehensive database on burst recurrence time in a
time frame coincident with the RXTE data is the BeppoSAX-WFC database,
because the exposure time is large (1 to 5 Msec per source).  This
encompasses 2300 X-ray bursts from 54 sources (e.g., in 't Zand et al.
2004). We have used this database to determine the average burst
recurrence time, simply by dividing the total exposure time per source
through the number of bursts detected from that source. Some sources
were never seen bursting with the WFC because the bursts are probably
all below the detection threshold (4U 1708-40, XTE J1710-281, 4U
1746-37, XTE J1759-220, UW Crb and Cyg X-2) and we refrain from giving
numbers on burst recurrence time. The numbers for the other sources
are provided in Table~\ref{burstlist}. These are based on total
exposures for all observations that the sources were sufficiently
close to the optical axis that the sensitivity was high enough to
detect typical bursts for each source.  The minimum required detector
area ranges between 5 and 40\% of the optimum on-axis case. Similar
percentages of observation time were excluded (i.e., with too far
off-axis positions).  The 68\%-confidence errors are based on Poisson
statistics for the counted number of bursts which represents a worst
case because bursts commonly do not occur randomly but quasi
periodically. The derived burst recurrence times generally follow the
trend with accretion rate as described above.

\section{Results: six new candidate UCXBs}
\label{results}

Excluding 2 clear cases of a high inclination angle and, therefore,
large cos$i$ correction (EXO 0748-676 and XTE J1710-281), there are 8
persistent X-ray bursters with luminosities below $\approx$2\% of
Eddington that are not identified yet as UCXBs, see
Table~\ref{tabucxb}. All of these are infrequent bursters (i.e. with
recurrence times in excess of a few days), which is consistent with
low accretion rates. We propose that these are good candidates for
being UCXBs. Two other cases were already previously identified on the
same grounds (in 't Zand et al. 2005a). The remaining six cases are in
order of right ascension:

\subsection{SAX J1712.6-3739}

This source was discovered in 1999 (in 't Zand et al. 1999; Cocchi et
al. 1999 and 2001) and since then is a persistent source (in 't Zand
et al. 2002b), if not earlier: there is a ROSAT All-Sky Survey
detection of a source just 0\farcm6 from the SAX position: 1RXS
J171237.1-373834 (in 't Zand et al. 1999). In Fig.~\ref{fig1712} is
shown the most detailed long-term light curve obtained thus far, with
the PCA bulge scan program. During these two years the source is
continuously active, apparently in two states: a slowly changing
state, and a quicker one. The all-time high in the flux is 230
c~s$^{-1}$~PCU$^{-1}$ or 1.6$\times10^{-9}$~\ecs\ or 3\% of the
burst-measured Eddington limit.

\begin{figure}[!t]
\centering
\includegraphics[width=0.99\columnwidth,angle=0]{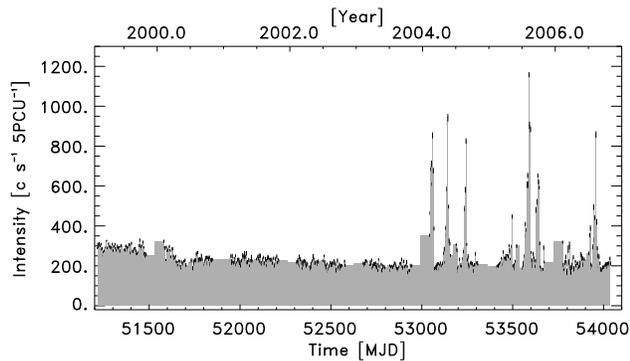}
\caption{The 2-12 keV light curve of 4U 1722-30 as measured
between February 1999 and November 2006 through the PCA bulge scan program.
\label{figterzan2}}
\end{figure}

In the 4.1~Ms large BeppoSAX-WFC database just a single photospheric
radius-expansion burst was detected with a derived distance of 6--8
kpc (Cocchi et al. 2001).  Recently, Chelovekov et al. (2006) reported
two further burst detections with INTEGRAL-IBIS in 2003-2004. All had
the same peak flux.

No optical counterpart has been identified yet within the 13\arcsec\
(1$\sigma$) ROSAT error circle radius. The X-ray absorption column
density of $N_{\rm H}=1.3\times10^{22}$ cm$^{-2}$ (Cocchi et al. 2001;
Dickey \& Lockman 1990) suggests a visual extinction of $A_{\rm
V}=7.3$. Together with the 7 kpc distance this brings the expected
visual magnitude to $\approx26$ for an ultracompact and $\approx22$
for a non-ultracompact binary. Refinement of the error circle through
Chandra and optical follow-up may bring confirmation of the UCXB
nature.

\subsection{4U 1722-30}

This is the bright LMXB in the globular cluster Terzan 2. It was first
detected 35 years ago with Uhuru. A 7-yr long X-ray light curve is
presented in Fig.~\ref{figterzan2}.  It shows the same behavior as for
SAX J1712.6-3739 with a slow/faint and quick/bright component,
although only in a limited time interval when the slow component is
faintest in 2-10 keV. Figure~\ref{figterzan2B} shows the light curve
zoomed in on four intervals of flaring activity. The shortest typical
time scale between flares is 50-100~d.

This system shows regular bursts with an average recurrence time of
2.5~d according to the BeppoSAX-WFC data archive which contains 24
burst detections (Cocchi et al. 2000b; Kuulkers et al. 2003).

\begin{figure}[!t]
\centering
\includegraphics[width=0.99\columnwidth,angle=0]{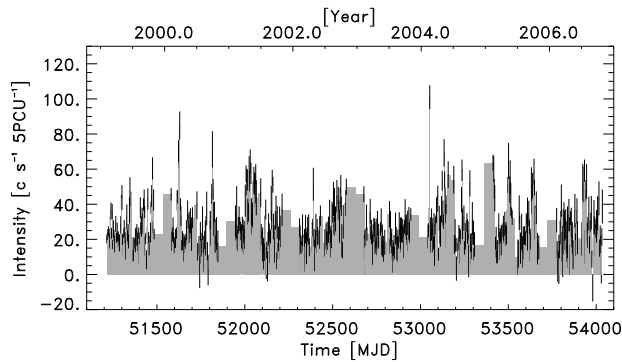}
\caption{The 2-12 keV light curve of 1RXS J172525.2-325717 as measured
between February 1999 and November 2006 through the PCA bulge scan
program.
\label{fig1725}}
\end{figure}

The reddening to Terzan 2 is $E(B-V)=1.57$ (Ortolani et al. 1997),
implying a visual extinction of $A_V=4.8$ (for $R=3.1$). The distance
of 9.5~kpc (e.g., Kuulkers et al. 2003) brings the expected apparent
visual magnitude to $\approx24$ for an ultracompact and $\approx20$
for a non-ultracompact binary. However, the source is located in the
core of the cluster where source confusion may be too much of an issue
for optical identification, even with a Chandra-determined position.

\subsection{1RXS J172525.2-325717}

1RXS J172525.2-325717 is a persistent source that is continuously
detected in the PCA bulge scans at a low flux, see
Fig.~\ref{fig1725}. There has not been a detailed study of its
persistent radiation yet. It is also known as IGR J17254-3257 (see
also Stephen et al.  2005; Walter et al. 2004). The only type-I X-ray
burst was discovered in 17 February 2004 3--30 keV data of the JEM-X
camera on INTEGRAL (Brandt et al. 2006). The burst had a rise time of
less than 5 s and an e-folding decay time of 15 s. The peak flux was
0.8 Crab units. For a 2.5~keV spectrum this would translate to roughly
a bolometric flux of 2.3$\times10^{-8}$~\ecs.

\subsection{SLX 1735-269}

\begin{figure}[!t]
\centering
\includegraphics[width=0.99\columnwidth,angle=0]{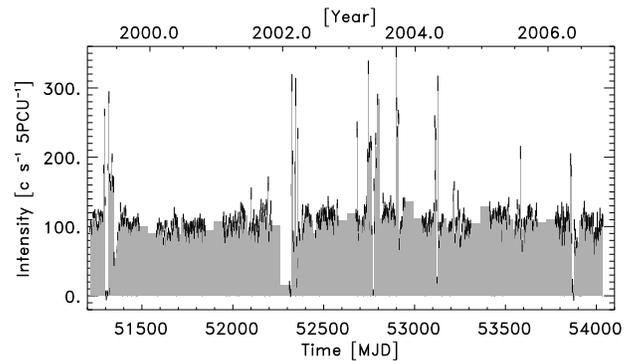}
\caption{The 2-12 keV light curve of SLX 1735-269 as measured between
February 1999 and November 2006 through the PCA bulge scan program.
\label{figslx1735}}
\end{figure}

\begin{figure*}[!t]
\centering
\includegraphics[width=1.99\columnwidth]{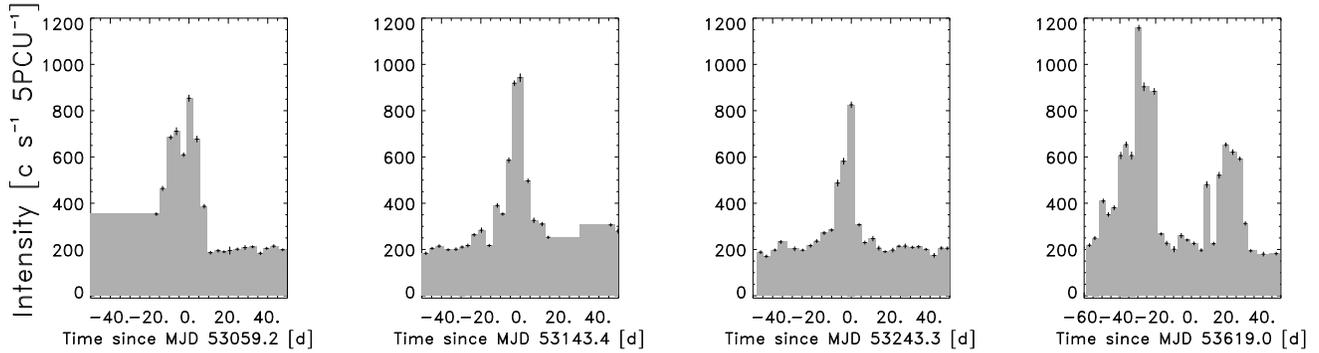}
\caption{The lightcurve of 4U 1722-30 (Fig.~\ref{figterzan2}), zoomed
in on the flares.
\label{figterzan2B}}
\end{figure*}

\begin{figure*}[!t]
\centering
\includegraphics[width=1.99\columnwidth]{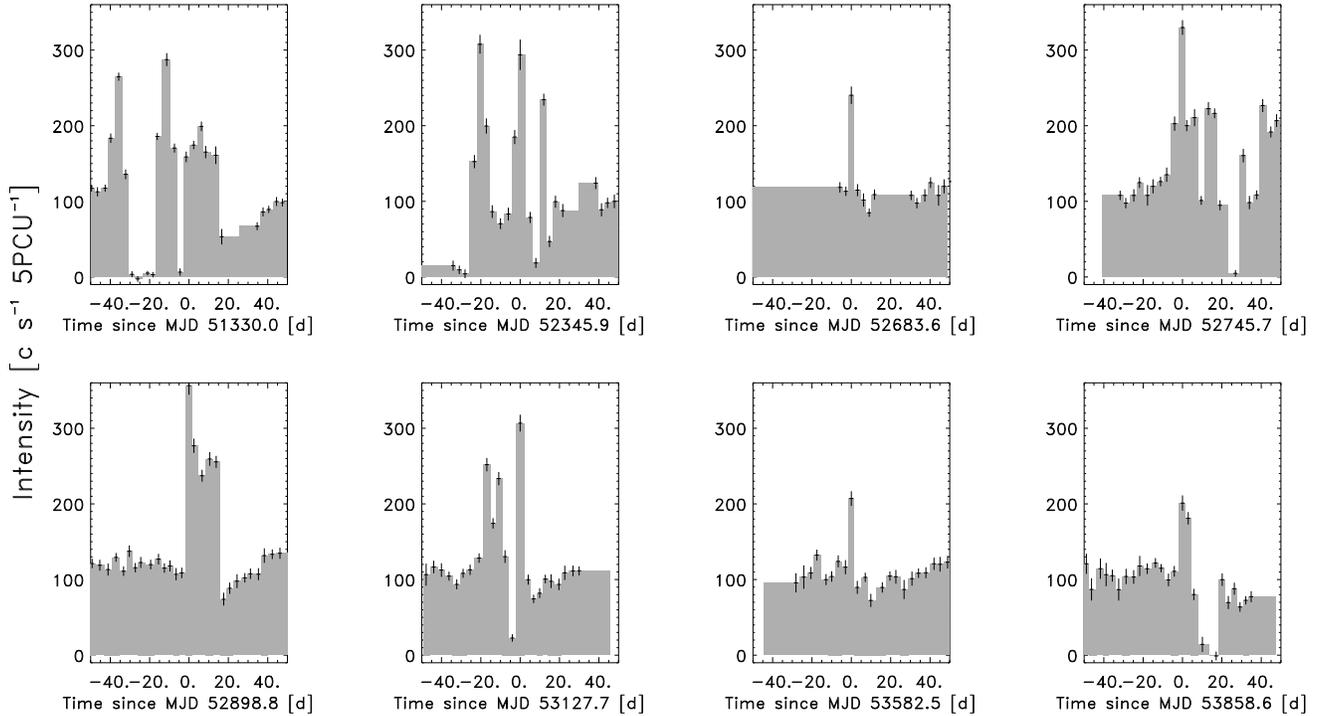}
\caption{The lightcurve of SLX 1735-269 (Fig.~\ref{figslx1735}),
zoomed in on a 100-d interval around the flares.
\label{figslx1735B}}
\end{figure*}

\begin{figure}[!t]
\centering
\includegraphics[width=0.99\columnwidth]{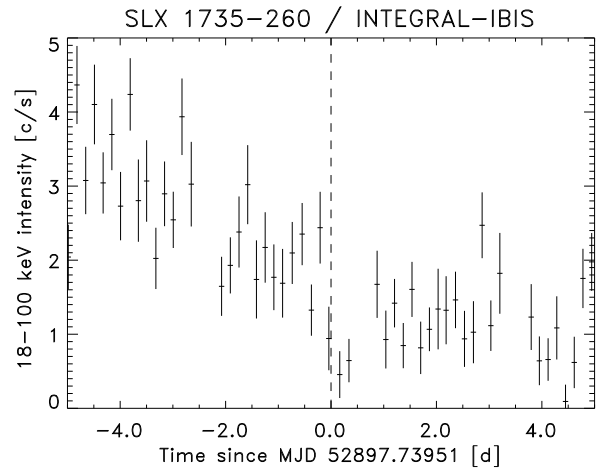}
\caption{The 18-100 keV light curve of SLX 1735-269 as measured around
the occurrence of the very long X-ray burst on Sep. 15, 2003 (as
indicated by the dashed line; Molkov et al. 2005) at a time resolution
of 4 INTEGRAL science windows (2.5~hr).
\label{slx1735integral}}
\end{figure}

SLX 1735-369 was discovered in 1985 data taken with the X-ray telecope
on Spacelab-2 (Skinner et al. 1987). The source was also detected in
1979 Einstein data by Elvis et al. (1992). It appears to be
continuously on. Figure~\ref{figslx1735} shows the X-ray light curve,
again exhibiting the same bimodal behavior as the previous two
sources. On top of that, remarkably the flux dips to zero sometimes
after a flare, see Fig.~\ref{figslx1735B}.

Discovered as an X-ray burster a decade ago through a single short
burst with BeppoSAX-WFC (Bazzano et al. 1997), SLX 1735-269 revealed 6
bursts with INTEGRAL (Molkov et al. 2005), one of them being of an
extremely energetic and long kind although probably not a
carbon-fueled superburst (see Fig.~\ref{figbursts}). This burst
occurred at the start of a brief outburst episode. This prompts the
question: is the outburst a result of the energetic burst or vice
versa? We investigated the INTEGRAL-IBIS data on the persistent
emission at somewhat higher resolution (2.5~hr resolution instead of
3~d; see Fig.~\ref{slx1735integral}) and find that the transition to
the high accretion rate state started 2~d prior to the
burst. Therefore, the long burst probably ignited as a result of an
increased accretion rate and not vice versa.

On June 20, 2005, another energetic burst was detected with HETE-2
(Suzuki \& Kawai 2005). Due to absence of PCA data it is not possible
to verify another association with an outburst.

The absorption column towards SLX 1735-269 is
$1.7\times10^{22}$~cm$^{-2}$ (Wilson et al. 2003), equivalent to $A_V
= 9.5$. For a distance of 6.5~kpc (Molkov et al. 2005) the expected
visual magntitude is $\approx27$ for an ultracompact and $\approx23$
for a non-ultracompact binary. Deep optical follow-up of the Chandra
position (Wilson et al. 2003) may bring confirmation of the UCXB
nature.

\subsection{SLX 1737-282}

SLX 1737-282 was first detected in 1985 with the Spacelab-2 mission
(Skinner et al. 1987) and was seen to radiate at the same flux level 5
times over the years (in 't Zand et al. 2002a). One burst was ever
detected from this source despite large exposure times and the
duration of the burst was extreme (the e-folding decay time 682 s is
the longest ever measured apart from superbursts). The PCA bulge scan
light curve is flat as well untill at least May 2005 after which it
suffered source confusion from nearby transients. We performed an
observation with the Swift X-Ray Telescope (XRT; Burrows et al. 2005)
on Oct. 10, 2006, and find a 0.5-10 keV unabsorbed flux of
$(1.36\pm0.15)\times10^{-10}$~\ecs, when fitting an absorbed power law
in the 1--8 keV band ($N_{\rm H}=(1.9\pm0.2)\times10^{22}$~cm$^{-2}$,
photon index 2.08$\pm0.15$ and $\chi^2_{\rm red}=0.6$ with 40 degrees
of freedom), which is consistent with all previous measurements since
discovery (cf, in 't Zand et al. 2002).

No optical counterpart has been identified yet within the 8\farcs\ 3
(90\% confidence) ROSAT error circle radius (in 't Zand et
al. 2003). The X-ray absorption column density of $N_{\rm
H}=1.9\times10^{22}$ cm$^{-2}$ suggests a visual extinction of $A_{\rm
V}=10.6$. Combined with the 5--8 kpc distance this brings the expected
visual magnitude to $\approx29$ for an ultracompact and $\approx25$
for a non-ultracompact binary. Confirmation of the UCXB nature through
optical follow up may prove cumbersome.

\subsection{SLX 1744-299}

\begin{figure}[!t]
\centering
\includegraphics[width=0.99\columnwidth,angle=0]{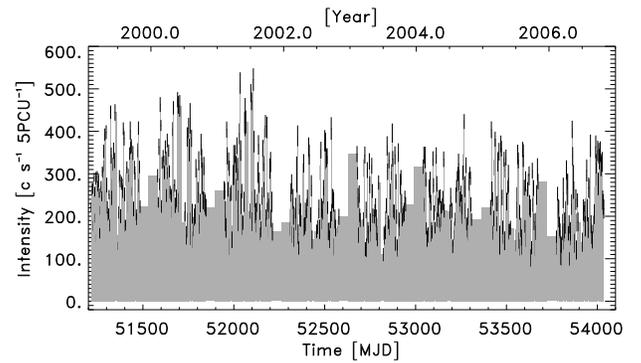}
\caption{The 2-12 keV light curve of SLX 1744-299 and 1744-300 combined
as measured between February 1999 and November 2006 through the PCA bulge scan program.
\label{figslx1744}}
\end{figure}

The nature of SLX 1744-299 (2\farcm8 from another X-ray burster SLX
1744-300; Skinner et al. 1987 and 1990) was established by Pavlinsky
et al. (1994). They detected one long X-ray burst of several hundred
seconds. Due to the small angular separation between both sources, all
non-focusing X-ray telescopes are only able to measure the combined
flux. A 2004 observation with XMM-Newton resolved both sources and
found a 2.8/1.0 flux ratio in the 0.5-10 keV band between SLX 1744-299
and SLX 1744-300 (Mori et al. 2005). This is the only such
measurement. We applied this flux ratio to the ASM and PCA data in
Table~\ref{burstlist}. The applicability of such a ratio is limited,
since the PCA bulge scan light curve shows considerable variability by
roughly a factor of two (Fig.~\ref{figslx1744}).

BeppoSAX-WFC detected 48 bursts from both sources. Three of these are
relatively long and can be identified with SLX 1744-299. The other 45
bursts are short and twice as faint.  This is consistent with archival
burst measurements which consistently reveal long and relatively
bright bursts from SLX 1744-299 and short and faint ones from SLX
1744-300. The longevity and slow recurrence of bursts from SLX
1744-299 are consistent with a UCXB nature.

\section{Confirmation of low $\dot{M}$ through a hard X-ray color}
\label{integral}

Bird et al. (2006) present a catalog of 209 soft $\gamma$-ray sources
detected with IBIS on INTEGRAL between February 2003 and June 2004.
The catalog includes the average fluxes in 2 photon energy bands:
20--40 and 40--100 keV. We made a selection of all 31 persistent X-ray
bursters in this list and ranked them according to their 40-100/20-40
keV hardness ratio, see Table~\ref{integraltable}. It includes 8
previously identified UCXBs and 3 new ones proposed here. The result
is fascinating. Almost all of the UCXBs have top rankings,
constituting the hardest persistent X-ray bursters. There are 2
exceptions: 4U 1820--30 and 4U 1915--05.

The hard nature of the UCXBs is consistent with the low mass accretion
rate inferred above (for a recent review of the spectral behavior
versus mass accretion rate in NS LMXBs, see van der Klis 2006; see
also Paizis et al. 2006). Therefore, this ranking is a confirmation of
the low mass accretion rate. The two exceptions are also consistent
with this. 4U 1820--30 has an accretion rate between 4 and
10$\times10^{-9}$ M$_\odot$~yr$^{-1}$ (Cumming 2003; note also that
$P_{\rm orb}$ [Stella et al. 1987] is the shortest of all). 4U 1915-05
is a high-inclination system while none of the other UCXBs in
Table~\ref{integraltable} are and, therefore, probably a substantial
multiplication factor is needed to correct for anisotropy of the
persistent radiation and the system likely has a mass accretion rate
of at least a few percent of Eddington. This may imply that it is
similar to 4U~1820-303 which may be supported by the fact that
4U~1915-05 is the only UCXB to show a helium line in the optical
spectrum (Nelemans et al. 2006b).

%Among the hard sources there also 2 with confirmed non-ultracompact
%orbital periods (IGR J17597-2201 and XTE J1710-281) and the reason
%that those are so high are because they are eclipsing systems and we
%possibly mostly only see the ADC source(?)

\begin{table}
\caption[]{Selection of all persistent X-ray bursters (ergo, LMXBs
with a NS accretor) from the 2nd IBIS/ISGRI catalog (Bird et
al. 2006), ranked according to increasing 40-100/20-40 keV hardness
ratio.\label{integraltable}}
\begin{center}
\begin{tabular}{llrl}
\hline\hline
Name            & Type       &  40--100 & UCXB?\\
                & of src.$^1$&  20--40  & \\
                &            &  flux & \\
                &            & ratio & \\
\hline
4U 1735-444     & B,A   &   0.04$\pm$0.23 &	\\
GX 17+2         & B,Z   &   0.05$\pm$0.09 &	\\
GX 3+1          & B,A   &   0.06$\pm$0.25 &	\\
4U 1820-303     & G,B,A &   0.07$\pm$0.13 & y \\
Ser X-1         & B     &$<$0.09$\pm$0.02 &	\\
Cyg X-2         & B,Z   &$<$0.10$\pm$0.02 &	\\
GX 13+1         & B,A   &   0.29$\pm$0.06 &	\\
GX 354-0        & B,A   &   0.38$\pm$0.01 &	\\
4U 1746-370     & G,B,A &   0.40$\pm$0.16 &	\\
4U 1915-05      & B,D   &   0.42$\pm$0.08 & y\\
4U 1636-536     & B,A   &   0.55$\pm$0.03 &	\\
1A 1742-294     & B     &   0.56$\pm$0.03 &	\\
4U 1254-690     & B,D   &$<$0.56$\pm$0.12 &	\\
4U 1705-440     & B,A   &   0.60$\pm$0.02 &	\\
4U 1702-429     & B,A   &   0.62$\pm$0.03 &	\\
4U 1323-62      & B,D   &   0.63$\pm$0.17 &	\\
SLX 1744-299    & B     &   0.65$\pm$0.04 & new \\
4U 1708-40      & B     &$<$0.67$\pm$0.17 &	\\
2S 0918-549     & B     &   0.72$\pm$0.16 & y\\
GS 1826-24      & B     &   0.80$\pm$0.01 &	\\
SLX 1735-269    & B     &   0.81$\pm$0.03 & new\\
4U 1850-087     & G,B   &   0.83$\pm$0.09 & y\\
4U 1722-30      & G,B,A &   0.88$\pm$0.02 & new\\
4U 0614+091     & B,A   &   0.92$\pm$0.07 & y\\
XTE J1759-220   & B,D   &   0.93$\pm$0.05 & \\
4U 1812-12      & B     &   0.99$\pm$0.01 & y\\
1A 1246-588     & B     &   1.00$\pm$0.25 & y\\
SLX 1737-282    & B     &   1.03$\pm$0.07 & new\\
4U 1705-32      & B     &   1.03$\pm$0.12 & y\\
1RXS J172525.2$^2$&B& 1.11$\pm$0.15 & new \\
XTE J1710-281   & B,E   &   1.37$\pm$0.10 & \\
\hline\hline
\end{tabular}
\end{center}

\noindent
$^1$G -- in globular cluster; B -- X-ray burster; A -- Atoll source; Z
-- Z source; D -- dipper; E -- eclipser; $^2$full name 1RXS
J172525.2-325717
\end{table}

\section{Discussion}
\label{discussion}

In summary, we have ranked the estimated average mass accretion rate
for all 40 X-ray bursters that have been active for at least 5 years
and identified 16 cases with rates below $\approx$2\% of the Eddington
limit that do not have a high inclination angle. These include 9 UCXBs
previously established on the basis of $P_{\rm orb}$ measurements or
low $L_{\rm opt}/L_{\rm X}$ values. We propose that the remaining
eight cases are UCXBs as well, 2 of which have been proposed already
in a preliminary study (in 't Zand et al. 2005a).

The correspondence between the low persistent flux and a low mass
accretion rate is supported by 2 burst characteristics: 1) the
recurrence times are relatively long: from 2.5 days (e.g., 4U 1722-30)
up to at least 2 weeks (e.g., 1RXS J171824.2-402934, see in 't Zand et
al. 2005a; see Table~\ref{burstlist}); 2) occasionally very long
bursts are observed, falling just short of the superburst regime with
e-folding decay times of up to 0.2 hr, which can be explained by a
longer fuel accumulation time implied by the lower accretion rate and
the cooler fuel temperature.

The long-term light curves of some newly identified UCXBs appear to
exhibit a peculiar bimodal behavior with a slowly varying component
(time scale hundreds of days) and a quickly varying component (time
scale a few days). One newly identified UCXB, SLX 1735-269, shows an
additional interesting feature in its light curve: an occasional
complete drop of the X-ray flux. Aside from these dips, variability of
the same magnitude and time scales has been observed in (presumable)
non-ultracompact X-ray binaries as well, for instance in the bright
sources GX 5-1, GX 9+1, GX 9+9, GX 340+0 and GX 349+2. The difference
lies in the fact that the duty cycle of the fast component is much
smaller in our UCXB candidates: the recurrence times of the flares in
the fast component are of order tens of days in UCXB candidates while
they are of order days in the bright non-ultracompact sources. It is
tempting to suggest that this is due to a difference in mass
ratio. For the UCXBs with pulsars, the mass ratio always is
$q=M_1/M_2<0.1$ for probable inclination angles far from 0$^{\rm o}$
(see references in Table~\ref{tabucxb}). It is well known that
mass-transfering binaries with $q\la0.3$ are susceptible to tidal
instabilities because the Kepler orbit around the accretor at which
there is a 3:1 resonance between the Kepler frequency and binary orbit
frequency then is inside the Roche lobe (Whitehurst 1988). This is
thought to possibly result in an eccentric disk precessing with
respect to the binary orbit which may modulate the accretion rate. In
CVs this mechanism is thought to give rise to the so-called
superoutbursts resulting from a combination of a thermal-viscous and a
tidal instability in the disk.  Perhaps in our systems only the tidal
instability is active and the thermal-viscous instability is absent
because the systems are persistent. The reason why SLX 1735-269
sometimes dies out completely is unclear. Obscuration by a warped disk
appears an attractive explanation. We note that this behavior is not
seen in all UCXBs. Possibly the effect is a sensitive function of
$\dot{M}$, as suggested by the light curve of 4U 1722-30
(Fig.~\ref{figterzan2}).  In conclusion, the peculiar light curve
features point out possibly interesting implications for accretion
disk theory.

A large, possibly dominant, fraction of LMXBs may be ultracompact. If
the 6 new UCXBs are valid, 18 out of the 40 persistent bursters are
ultracompact.  Possibly several more are ultracompact, because not all
remaining systems yet have optical counterparts through which an
ultracompact nature may be indicated; some could be similar to 4U
1820-303.  The fraction of ultracompact cases is 7/8 in globular
clusters (4U 1746-37 being the sole exception) and between 11/32 and
21/32 in the Galactic field (the upper limit being defined by the 11
LMXBs with measured non-ultracompact periods). The numbers are too
small to analyze differences between the globular cluster and Galactic
field populations in a meaningful way.

Ultimate verification of an ultracompact nature can only be done
through measurement of $P_{\rm orb}$. The most succesful (tentative)
measurements were done through optical photometry. Many of the
candidate UCXBs are located near the Galactic center, implying large
extinctions of at least 5 mag in $V$.  Combined with a distance
modulus of 14.5 (for a canonical 8~kpc distance) and an expected $M_V$
of about 4 (van Paradijs \& McClintock 1994), renders photometry
towards the infrared with 6~m class telescopes as the only possible
route.

\acknowledgement 

We are grateful to Andrew Cumming, Gijs Nelemans, Duncan Galloway,
Michael Truss, Jean-Pierre Lasota and Laurens Keek for very useful
discussions, , Lucien Kuiper and Peter den Hartog for help with
understanding INTEGRAL data, Sergey Molkov for providing the processed
INTEGRAL data on the burst from SLX 1735-269 and Frank Verbunt for
providing useful comments on an earlier version of the manuscript.  We
thank the Swift team for granting and preparing a target of
opportunity observation of SLX 1737-282 to clarify our flux
measurements with the PCA bulge program. This research has made use of
the RXTE/ASM archive provided by ASM teams at MIT and at the RXTE SOF
and GOF at NASA/GSFC, of the INTEGRAL data archive provided by the
INTEGRAL Science Data Center at the Observatoire de Gen\`{e}ve, and of
data from the Italian-Dutch BeppoSAX mission provided by the ASI
Science Data Center in Rome.  JZ acknowledges support from The
Netherlands Organization for Scientific Research (NWO).

\end{document}